\documentclass[12pt]{article}
\usepackage{amsmath}
\usepackage{amssymb}
\usepackage{amsfonts}

\usepackage[nomath]{stix}
\usepackage{setspace}
\usepackage[T1]{fontenc}
\usepackage{times}
\usepackage{palatino} 
\usepackage{lmodern}
\usepackage[latin1]{inputenc}
\usepackage{epsfig}
\usepackage[english]{babel}
\usepackage{color}
\usepackage{graphicx}
\usepackage{dcolumn}
\usepackage{moreverb}
\usepackage{cite}
\usepackage{auto-pst-pdf} 

\usepackage{arydshln}
\usepackage{cite}
\usepackage{braket}

\usepackage{dsfont}
\usepackage{nicefrac}
\usepackage{graphicx}
  \usepackage{stmaryrd}
\usepackage{epstopdf}

\usepackage{array}
\usepackage{esvect}
\usepackage{cite}

\usepackage{physics}
\usepackage{mathrsfs}
\usepackage{authblk}

\usepackage[all,cmtip]{xy}

\numberwithin{equation}{section}

\title{Special function models of indecomposable $sl(2)$ representations:
The Laguerre Case}
\author{S\'ebastien Bertrand$^{1}$, Ian Marquette$^2$, Willard Miller Jr $^{3}$, Sarah Post$^{1}$   \\
{\small $^1$ Department of Mathematics, University of Hawaii, Honolulu, HI, 96822, USA} \\ 
{\small $^2$ School of Mathematics and Physics, The University of Queensland, }\\ 
{\small Brisbane, QLD 4072, Australia }\\ 
{\small $^3$  School of Mathematics, University of Minnesota, Mineapolis, MN, 55455, USA} 

}

\date{ }
\begin{document}
\baselineskip=20pt plus 1pt minus 1pt
%%%%%%%%%%%%%%%%%%%%%%%%%%%%%%%%%%%%%%%%%%%%%%%%%%%%%%%%%%
\maketitle

\begin{abstract}
 In this paper, we point out connections between certain types of indecomposable representations of $sl(2)$ and generalizations of well-known orthogonal polynomials. Those representations take the form of infinite dimensional chains of weight or generalised weight spaces, for which the Cartan generator acts in a diagonal way or via Jordan blocks. The other generators of the Lie algebras $sl(2)$ act as raising and lowering operators but are now allowed to relate the different chains as well. In addition, we construct generating functions, we calculate the action of the Casimir invariant and present relations to systems of non-homogeneous second-order coupled differential equations. We present different properties as higher-order linear differential equations for building blocks taking the form of one variable polynomials. We also present insight into the zeroes and recurrence relations.
\end{abstract}

%The modules of those representations take the form of two variables functions involving certain polynomials as building blocks.

%Lie algebras have been associated with various realizations as differential operators and corresponding special functions. The Lie algebra $sl(2)$ have been studied extensively from point of view of differential operators realizations. Those works have been stimulated by applications in various areas of mathematical physics and the use of Lie algebras to describes various symmetries of Hamiltonians in quantum physics. Most of those links concerned irreducible representations.

\section{Introduction}

Lie algebras, their realizations and connections with special functions have attracted a lot of attention in the area of mathematical physics since the end of the 60s \cite{ism67, mil68, mil70,mil72,mil73a,mil73b,kal80}. This has been inspired partly by the fruitful connection of special functions and Lie algebras with quantum mechanical problem \cite{frad65, bar65, win67}. In this context, the generators of the Lie algebras can be integrals of motion, ladder operators or other physically relevant operators. Via the notion of hidden symmetry underlying exact and quasi-exact solvability, the $sl(2)$ Lie algebra was connected to various models  \cite{tur88}. The realizations related to other higher-rank Lie algebras such as $sl(n)$ and $so(n)$ have lead to profound connection with generalized hypergeometric, Lauricella and Horn functions (see e.g. \cite{mil68}). Those special functions have appeared in different contexts of mathematical physics. 

The cases of the Lie algebras $sl(2)$ with one and two variables realizations have been intensively described in regard of unitary and irreducible representations of $sl(2)$. This rank-1 algebras has been related to different orthogonal polynomials, in particular using first but also second-order differential operators \cite{koo78, bas82, bas85, koo86, bas87, bas89, bac90}. An extensive description of Lie algebras and their realizations, and in particular from point of view of the Casimir invariant was developed by Vilenkin and Klimyk \cite{vil91}. Some connections with wider types of representation beyond irreducible and unitary was pointed out \cite{how92}, in regard of indecomposable representations \cite{hum08, maz09,su93} and quantum field theory \cite{ras06}. 

Among the indecomposable representations, there are different cases that consist in combinations of finite and infinite dimensional representations glued together via the action of the Cartan or other generators and forming interconnected chains. The paper intends to generalize certain indecomposable representations in context of $sl(2)$ with arbitrary number of chains from the perspective of differential operators that were obtained in context of the Laguerre case\cite{mil68}. This will lead to constructing various polynomials where the raising generators of $sl(2)$ will act almost as usual ladder operators but allowing interaction between the different chains. The papers will also establish the connection with systems of coupled nonhomogeneous second-order differential equations. 

Standard literature on the theory of ODEs and special functions has mostly looked at uncoupled homogeneous ODEs in the hypergeometric class and their generalisation as the Heun equation \cite{inc56,koe02}. The study of systems of non-homogeneous differential equations, their solutions and connections to special functions have been much less studied. Some cases related to resonant problem \cite{bab67,mil70a, zac76, bac86,bac01} were studied. However, there is no systematic search for solution of those systems of ODEs. This paper allows to discuss some classes of such systems of ODEs, that naturally extend the ODE of the Laguerre polynomials, and with an underlying Lie algebras such as sl(2) and related indecomposable representations. This makes those polynomials and special functions of interest for applications in mathematical physics. Here the connection between indecomposable representations and $sl(2)$ provides a new perspective on this topic.

The papers is organised as follow. In Section 2, we set the foundation and notation of the manuscript while reviewing properties of the (generalized) Laguerre polynomials. In Section 3 we present the case of indecomposable with chains where T, the Cartan generator, act as a Jordan block. We also present explicitly the polynomials solving the systems of nonhomogenous coupled ODEs. We give extra details on the 2 and 3 chains problems. In particular we present second-order differential-recurrence relations, generating functions and formula in terms of linear combination of Laguerre polynomials. In Section 4 we characterise the case of Laguerre polynomials where T is diagonalizable (but the raising/lowering operators are not). In Section 5 we provide further insight into some of the building block taking the form of new polynomials. We present the third and fourth order ODE that they satisfy, further recurrence relations, related family of polynomial and bi-orthogonality type relations and insight into the zeroes of those polynomials. In Section 6, we discuss how those are in fact connected with certain representations of $sl(2)$ where the infinite dimensional representation is obtained via induced representation of a finite representation of the Borel subalgebra.

\section{Preliminaries}

The motivation for the treatment of indecomposable representations presented here takes origin in the large body of work on $sl(2)$ and what we would refer as one-chain representations, and studied in the monograph \cite{mil68} in which the differential recurrence relations for hypergeometric, confluent hypergeometric and Bessel functions were shown to be expressible as representations of the three four-dimensional Lie algebras $sl(2) \oplus (E)$, the Heisenberg algebra, and $e(2)\oplus (E)$. The differential recurrence relations for these three types of hypergeometric functions are linear first-order ordinary differential equations for the functions $F_n(z)$. The idea is to add a new variable $t$ and to replace the ordinary differential equations by partial differential equations in $z,t$ such that in the new system the basis functions are  $F_n(z)t^n$. If this is done correctly the new recurrence operators will close to generate one of the three Lie algebras listed above. The advantage of the addition of a 2nd variable is that the tools of Lie theory become available and we can find new properties of these functions, such as generating functions with relative ease. 

In the monograph \cite{mil68} these realizations were used to classify the finite and infinite dimensional bounded above or below irreducible representatives of  $sl(2)\oplus(E)$, the Heisenberg algebra, and $e(2)\oplus(E)$, most of which do not extend to global representations of the corresponding groups. However, local Lie group theory still applies and can be used to derive additional theorems and generating functions where global group representations no longer apply.

In this paper we shall employ the realization of type B ($sl(2)$, Laguerre polynomials) of the differential recurrence relations employed in \cite{mil68}, but now used to find realizations of indecomposable representations of $sl(2)$.
\newline
\newline
If we restrict ourselves to one-chain in the type B, the (generalized) Laguerre  are known to possess raising and lowering operators forming an $sl(2)$ algebra, that is
\begin{equation}
[E^-,E^+]=2T,\qquad [T,E^{\pm}]=\pm E^\pm,\label{sl2-com}
\end{equation}
where $E^+$ and $E^-$ are respectively the raising and lowering operators. The $\mathfrak{sl}_2$ algebra possess the following Casimir element:
\begin{equation}
K=T^2+T+E^-E^+.\label{sl2-Casi}
\end{equation}

\subsection{Generalized Laguerre polynomials}
For the generalized Laguerre polynomials, which can be defined using a generalized hypergeometric function,
\begin{equation}
L_n^{(\alpha)}(z)=\dfrac{\Gamma(\alpha+n+1)}{\Gamma(\alpha+1)}\hspace{2mm}F_{\hspace{-3.5mm}1\hspace{2mm}1}\left(\begin{array}{c}
-n\\
\alpha+1
\end{array},z\right),
\end{equation}
the raising and lowering operators can be expressed as follow
\begin{equation}
E^+=z\dfrac{d}{dz}+n+1+\alpha-z,\qquad E^-=z\dfrac{d}{dz}-n\label{RL-Lag}
\end{equation}
with the additional operator
\begin{equation}
T=n+\dfrac{1+\alpha}{2}.
\end{equation}
The raising and lowering operators satisfy the differential recurrence relations
\begin{equation}
E^+L_n^{(\alpha)}(z)=(n+1)L_{n+1}^{(\alpha)}(z),\qquad E^-L_n^{(\alpha)}(z)=-(\alpha+n)L_{n-1}^{(\alpha)}(z)
\end{equation}
and they satisfy the $sl(2)$ commutation relations (\ref{sl2-com}) together with $T$.
By combining these two recurrence relations, we can recover the differential equation
\begin{equation}
\left(z\dfrac{d^2}{dz^2}+(1+\alpha-z)\dfrac{d}{dz}+n\right)L_n^{(\alpha)}(z)=0.\label{diff-Lag}
\end{equation}
Using a linear combination of the raising/lowering operators and their actions, we can recreate the recurrence equation
\begin{equation}
(n+1)L_{n+1}^{(\alpha)}(z)=(2n+1+\alpha-z)L_{n}^{(\alpha)}(z)-(n+\alpha)L_{n-1}^{(\alpha)}(z).
\end{equation}
The generating function for the generalized Laguerre polynomials is given by
\begin{equation}
(1-t)^{-(\alpha+1)}\exp\left(\dfrac{-zt}{1-t}\right)=\sum_{n=0}^\infty L_n^{(\alpha)}(z)t^n.
\end{equation}

%\subsection{Hypergeometric polynomials}
%To be added...

\subsection{Many chains of polynomials}
The realizations of the operators $E^\pm$ and $T$ for the polynomials (or one chain) can be adapted to generate two or more chains of new polynomials that are linked to the previous chain. That is we introduce a new variable $t$ to allow a higher number of chains. For convenience, we will construct and orient the possible states on the chains as follow:
\begin{equation*}
\begin{array}{ccccc}
\vert 0,N\rangle & \vert 1,N\rangle & ... & \vert n,N\rangle & ...\\
\vdots & \vdots & & \vdots & \\
\vert 0,\beta\rangle & \vert 1,\beta\rangle & ... & \vert n,\beta\rangle & ... \\
\vdots & \vdots & & \vdots & \\
\vert 0,2\rangle & \vert 1,2\rangle & ... & \vert n,2\rangle & ... \\
\vert 0,1\rangle & \vert 1,1\rangle & ... & \vert n,1\rangle & ... 
\end{array}
\end{equation*}
where $N$ is the number of chains considered. The states on the first chain, $\vert n, N\rangle$ for $n\in\mathbb{N}$, are linked with the original polynomials, i.e. the auxiliary variable $t$ to the power $n$ times the generalized Laguerre polynomials
\begin{equation}
\vert n, N\rangle = L_n^{(\alpha)}(z)t^n.
\end{equation}
%and for the Hypergeometric polynomials
%\begin{equation}
%\vert n, N\rangle = H_n^{(bc)}(z)t^n,
%\end{equation}
The $\mathfrak{sl}_2$ elements can act centrally and map to the closest neighbours, cardinal and ordinal directions. The cardinal direction includes north, south, east and west. The ordinal includes the northeast, southeast, southwest and northwest. Each elements (excluding the top and bottom chains and the initial states on the right) have 8 nearest neighbors. We assume the existence of a highest weight going north, i.e. the original polynomials, and a lowest weight going west. (Depending on the case, the lowest weight in the West direction may not be the state $\vert 0,\beta\rangle$.) We will consider infinite-dimensional chains, that it there will be no highest weight in the East direction.

\section{Laguerre polynomials with non-central operators}
In this case, we will consider that none of the operators act centrally. One representation of such an algebra takes the form
\begin{eqnarray}
T\vert n,\beta\rangle &=& \left(n+\dfrac{\alpha+1}{2}\right)\vert n,\beta\rangle+\vert n,\beta+1\rangle,\\
E^+\vert n,\beta\rangle &=& (n+1)\vert n+1,\beta\rangle,\\
E^-\vert n,\beta\rangle &=& -(n+\alpha)\vert n-1,\beta\rangle-2\vert n-1,\beta+1\rangle,
\end{eqnarray}
i.e. $T$ acts centrally and north, $E^+$ acts east and $E^-$ acts west and north-west. It is important to note that on the highest chain, all the operators act like the operators in equation (\ref{RL-Lag}). Let us denote the vanishing of the north coefficients on the highest chain. A realization is of this representation is given by
\begin{eqnarray}
T&=&t\partial_t+\dfrac{1+\alpha}{2},\\
E^+&=&t(z\partial_z+t\partial_t+(1+\alpha-z)),\label{real-Lag}\\
E^-&=&\dfrac{z\partial_z-t\partial_t}{t},
\end{eqnarray}
which also satisfy the $sl(2)$ commutation relations (\ref{sl2-com}). 

To generate a new chain, one can use the equation
\begin{equation}
T\vert 0,N-1\rangle=\left(\dfrac{\alpha+1}{2}\right)\vert 0,N-1\rangle+\vert 0,N\rangle.
\end{equation}
Using the realization of $T$ and considering $\vert 0,N-1\rangle$ as an unknown function of $z$ and $t$, we obtain a linear differential equation that we can solve. We can then generate all the other elements of the chain by successively applying $E^+$ on each state $\vert n,N-1\rangle$. Finally, by checking the restriction of $sl(2)$, we can construct the whole states on the second chain. 

Once the second chain is constructed, we can reapply the same method to construct the third (or lower) chains. As a result, we can express all the states as
\begin{equation}
\vert n,\beta\rangle =\left(\sum_{j=0}^{N-\beta}\dfrac{\omega_{n,j}(z)\ln(zt)^{N-\beta-j}}{(N-\beta)!}\right)t^n,
\end{equation}
where $n$ is any natural number and $\beta\in \lbrace 1,...,N\rbrace$. The functions $\omega_{n,j}(z)$ are polynomials in $z$ satisfying the raising and lowering relations
\begin{eqnarray}
z\dfrac{d}{dz}\omega_{n,\beta}(z)+(n+\alpha+1-z)\omega_{n,\beta}(z)=(n+1)\omega_{n+1,\beta}(z)-2\omega_{n,\beta-1}(z),\\
z\dfrac{d}{dz}\omega_{n,\beta}(z)-n\omega_{n,\beta}(z)=-(\alpha+n)\omega_{n-1,\beta}(z)-2\omega_{n-1,\beta-1}(z).
\end{eqnarray}
By combining those two last equations, we can construct a linear second-order differential equation
\begin{equation}
z\dfrac{d^2}{dz^2}\omega_{n,\beta}(z)+(1+\alpha-z)\dfrac{d}{dz}\omega_{n,\beta}(z)+n\omega_{n,\beta}(z)=-2\dfrac{d}{dz}\omega_{n,\beta-1}(z),
\end{equation}
which is the generalized Laguerre differential equation (\ref{diff-Lag}) but with a non-homogeneous term. For the highest chain, we have by construction that $\omega_{n,N}(z)$ is the generalized Laguerre polynomials, i.e.
\begin{equation}
\omega_{n,N}(z)=L_{n}^{(\alpha)}(z).
\end{equation}
Note that any polynomial $\omega_{n,\beta}(z)$ where $n<0$ or $\beta<N$ is set to zero.

Using the representation above, the Casimir element $\mathcal{K}$ in equation (\ref{sl2-Casi}) does not act centrally, that is
\begin{equation}
\mathcal{K}\vert n,\beta\rangle=\dfrac{\alpha^2-1}{4}\vert n,\beta\rangle+\alpha\vert n,\beta+1\rangle+\vert n,\beta+2\rangle.
\end{equation}

By applying $E^+$ repetitively on a state, we have
\begin{equation}
(E^+)^k\vert n,\beta\rangle=\dfrac{(n+k)!}{n!}\vert n+k,\beta\rangle.
\end{equation}
Therefore, if we exponentiate the raising operator $E^+$ and apply it on a state, we obtain
\begin{equation}
\exp(uE^+)\vert n,\beta\rangle=\sum_{k=0}^\infty u^k\dfrac{(n+k)!}{n!k!}\vert n+k,\beta\rangle.
\end{equation}
If we take $n=0$, $\beta=N$ and $u=1$, then we obtain the equivalent of the generating function of the Laguerre polynomial
\begin{equation}
\exp(E^+)\vert 0,N\rangle=\sum_{k=0}^\infty L_k^{(\alpha)}(z)t^k=(1-t)^{-(\alpha+1)}\exp\left(\dfrac{-zt}{1-t}\right).\label{GF-Lag}
\end{equation}
In fact, applying the exponentiation of $E^+$ on any locally analytic function $f(z,t)$ (see \cite{mil68}) , we have that
\begin{equation}
\exp(uE^+)f(z,t)=(1-ut)^{-(\alpha+1)}\exp\left(\dfrac{-uzt}{1-ut}\right)f\left(\dfrac{z}{1-ut},\dfrac{t}{1-ut}\right).\label{GF-Lag-Jor}
\end{equation}

\subsection{The two-chain example}
In the case where we consider only two chains, i.e. $N=2$, let us define the vector
\begin{equation}
V_n=\left(\begin{array}{c}
\vert n, 2\rangle\\
~\\
\vert n, 1\rangle
\end{array}\right)=\left(\begin{array}{c}
L_n^{(\alpha)}(z)t^n\\
~\\
(L_n^{(\alpha)}(z)\ln(tz)+\omega_{n,1}(z))t^n
\end{array}\right).
\end{equation}
Then we have
\begin{eqnarray}
TV_n&=&\left(\begin{array}{cc}
n+\dfrac{\alpha+1}{2} & 0 \\
1 &n+\dfrac{\alpha+1}{2} 
\end{array}\right)V_n,\\
E^+V_n&=&\left(\begin{array}{cc}
n+1 & 0 \\
0 & n+1 
\end{array}\right)V_{n+1},\\
E^-V_n&=&\left(\begin{array}{cc}
-(n+\alpha) & 0 \\
-2 & -(n+\alpha)
\end{array}\right)V_{n-1}.
\end{eqnarray}
The Casimir $\mathcal{K}$ can be represented as the matrix satisfying
\begin{equation}
\mathcal{K}V_n=\left(\begin{array}{cc}
\dfrac{\alpha^2-1}{4} & 0 \\
\alpha & \dfrac{\alpha^2-1}{4}
\end{array}\right)V_n.
\end{equation}
The polynomial $\omega_{n,1}(z)$ can be solved completely in terms of Laguerre polynomials and $\omega_{0,1}$ which is a constant, that is
\begin{equation}
\omega_{n,1}(z)=\sigma_1L_n^{(\alpha)}(z)+\sum_{m=1}^n\dfrac{2}{m}L_{n-m}^{(\alpha)}(z),\qquad \mbox{where~~}\omega_{0,1}(z)=\sigma_1.
\end{equation}

Using the generation function in equation (\ref{GF-Lag-Jor}) with $u=1$ on the first state of the second chain, i.e. $\vert 0,1\rangle=\sigma_1+\ln(zt)$ and substracting the Laguerre generating function (\ref{GF-Lag}), we get a generating function for $\omega_{n,1}(z)$, i.e. 
\begin{equation}
(\sigma_1-2\ln(1-t))(1-t)^{-(\alpha+1)}\exp\left(\dfrac{-zt}{1-t}\right)=\sum_{k=0}^\infty\omega_{k,1}(z)t^k.
\end{equation}

\subsection{The three-chain example}
For the case with three chains, i.e. $N=3$, let us define the vector
\begin{equation}
V_n=\left(\begin{array}{c}
\vert n, 3\rangle\\
~\\
\vert n, 2\rangle\\
~\\
\vert n,1\rangle
\end{array}\right)=\left(\begin{array}{c}
L_n^{(\alpha)}(z)t^n\\
~\\
(L_n^{(\alpha)}(z)\ln(tz)+\omega_{n,1}(z))t^n\\
~\\
\left(\dfrac{L_n^{(\alpha)}(z)\ln(tz)^2}{2}+\omega_{n,1}(z)\ln(tz)+\omega_{n,2}\right)t^n
\end{array}\right).
\end{equation}
Then we have
\begin{eqnarray}
TV_n&=&\left(\begin{array}{ccc}
n+\dfrac{\alpha+1}{2} & 0 & 0 \\
1 &n+\dfrac{\alpha+1}{2}  & 0 \\
0 & 1 &n+\dfrac{\alpha+1}{2} 
\end{array}\right)V_n,\\
E^+V_n&=&\left(\begin{array}{ccc}
n+1 & 0 & 0 \\
0 & n+1  & 0 \\
0 & 0 & n+1
\end{array}\right)V_{n+1},\\
E^-V_n&=&\left(\begin{array}{ccc}
-(n+\alpha) & 0 & 0\\
-2 & -(n+\alpha) & 0 \\
0 & -2 & -(n+\alpha) \\
\end{array}\right)V_{n-1}.
\end{eqnarray}
The Casimir $\mathcal{K}$ can be represented as the matrix satisfying
\begin{equation}
\mathcal{K}V_n=\left(\begin{array}{cccc}
\dfrac{\alpha^2-1}{4} & 0  & 0\\
\alpha & \dfrac{\alpha^2-1}{4} & 0 \\
1 & \alpha & \dfrac{\alpha^2-1}{4}
\end{array}\right)V_n.
\end{equation}
The polynomial $\omega_{n,2}(z)$ can be solved completely in terms of Laguerre polynomials and $\omega_{0,2},\omega_{0,1}$ which are constants, that is
\begin{equation}
\omega_{n,2}(z)=\sigma_2L_n^{(\alpha)}(z)+\sum_{m=1}^n\left(\dfrac{2\sigma_1+\sum_{k=1}^{m-1}\dfrac{4}{k}}{m}L_{n-m}^{(\alpha)}(z)\right),\quad \mbox{where~~}\omega_{0,2}(z)=\sigma_2,\quad \omega_{0,1}(z)=\sigma_1
\end{equation}

or 

\begin{equation}
\omega_{n,2}(z)= \sigma_2 L_n^{(\alpha)}(z) + \sum_{k=0}^{n-1} \left[ \frac{2}{n-k} \sigma_1 + \frac{4(\gamma+\psi(0,n-k))}{n-k} \right] L_k^{(\alpha)}(z),
\end{equation}
where $\gamma$ is the Euler gamma and $\psi(n,z)$ the polygamma functions.

Using the generation function in equation (\ref{GF-Lag-Jor}) with $u=1$ on the first state of the third chain, i.e. $\vert 0,1\rangle=\sigma_2+\sigma_1\ln(zt)+\ln(zt)^2$, we obtain, after subtracting  the generating functions of $L_n^{(\alpha)}$ and $\omega_{n,1}$, that the generating function for $\omega_{n,2}$ is
\begin{equation}
\left(\sigma_2-2\sigma_1\ln(1-t)+2(\ln(1-t))^2\right)(1-t)^{-(\alpha+1)}\exp\left(\dfrac{-zt}{1-t}\right)=\sum_{k=0}^\infty\omega_{k,2}(z)t^k.
\end{equation}

\section{Laguerre polynomials with one central operator}
In the case where $T$ can be diagonalized, we can consider the actions of $T$, $E^+$ and $E^-$ as
\begin{eqnarray}
T\vert n,\beta\rangle &=& \left(n+\dfrac{\alpha+1}{2}\right)\vert n,\beta\rangle,\\
E^+\vert n,\beta\rangle &=& (n+1+\beta-N)\vert n+1,\beta\rangle,\\
E^-\vert n,\beta\rangle &=& (\beta-\alpha-n-N)\vert n-1,\beta\rangle+\vert n-1,\beta+1\rangle,
\end{eqnarray}
i.e. where $T$ acts centrally, $E^+$ acts east and $E^-$ acts west and north-west. A realization of this action is given the same realization as equation (\ref{real-Lag}), that is
\begin{eqnarray*}
T&=&t\partial_t+\dfrac{1+\alpha}{2},\\
E^+&=&t(z\partial_z+t\partial_t+(1+\alpha-z)),\\
E^-&=&\dfrac{z\partial_z-t\partial_t}{t}.
\end{eqnarray*}
Conversely to the previous case, we will consider that the lowest weight going west is not when $n=0$ but when $n=N-\beta$ for a state $\vert n,\beta\rangle$ for a system with $N$ state. For a finite number of chains, this arrangement of states takes the form of an infinite trapezoid.

For this representation, we cannot use $T$ to solve for the first state of the second chain, i.e. $\vert 1,N-1\rangle$, but we can use can use $E^-$. We have
\begin{equation}
E^-\vert 1,N-1\rangle=\vert 0,N\rangle.
\end{equation}
N.B. that the first chain is known and given by $\vert n,N\rangle=L_n^{\alpha}(z)t^n$. Once we the state $\vert 1,N-1\rangle$, we can generate all the other states using $E^+$. By checking all the commutation relations of $sl(2)$, and then generating $N$ chains, we obtain
\begin{equation}
\vert n,\beta\rangle = \omega_{n,\beta}(z)t^n,
\end{equation}
where $\omega_{n,\beta}(z)$ is polynomial in $z$. Those polynomials $\omega_{n,\beta}(z)$ satisfy the raising and lowering recurrence relations
\begin{eqnarray}
z\dfrac{d}{dz}\omega_{n,\beta}(z)+(n+\alpha+1-z)\omega_{n,\beta}(z)=(n-\beta+1)\omega_{n+1,\beta}(z),\\
z\dfrac{d}{dz}\omega_{n,\beta}(z)-n\omega_{n,\beta}=-(\alpha+n+\beta)\omega_{n-1,\beta}(z)+\omega_{n-1,\beta-1}(z).
\end{eqnarray}
By combining those two last equations, we can construct a linear second-order differential equation
\begin{equation}
z^2\dfrac{d^2}{dz^2}\omega_{n,\beta}(z)+z(1+\alpha-z)\dfrac{d}{dz}\omega_{n,\beta}(z)+(nz-\alpha\beta-\beta^2)\omega_{n,\beta}(z)=(n-\beta+1)\omega_{n,\beta-1}(z).
\end{equation}

The Casimir acts north and centrally, that is
\begin{equation}
\mathcal{K}\vert n,\beta\rangle=\dfrac{(2\beta+\alpha+1)(2\beta+\alpha-1)}{4}\vert n,\beta\rangle+(n+1-\beta)\vert n,\beta+1\rangle.
\end{equation}

By using $E^+$, it is possible to construct a generating function. We have
\begin{equation}
(E^+)^k\vert n,\beta\rangle=\dfrac{(n+k+\beta-N)!}{(n+\beta-N)!}\vert n+k,\beta\rangle.
\end{equation}
Hence,
\begin{equation}
\exp(uE^+)\vert n,\beta\rangle=\sum_{k=0}^\infty\dfrac{u^k}{k!}\dfrac{(n+k+\beta-N)!}{(n+\beta-N)!}\vert n+k,\beta\rangle.
\end{equation}
By taking $u=1$, $n=0$ and $\beta=N$ and using previous results, we get
\begin{equation}
(1-t)^{-(\alpha+1)}\exp\left(\dfrac{-zt}{1-t}\right)=\sum_{k=0}^\infty L_k^{(\alpha)}(z)t^k,
\end{equation}
which matches Laguerre's generating function.

\subsection{The two-chain example}
Let us define the vector
\begin{equation}
V_n=\left(\begin{array}{c}
\vert n,2\rangle \\
\\
\vert n,1\rangle
\end{array}\right)=\left(\begin{array}{c}
L_n^{(\alpha)}(z)t^n\\
\\
\omega_{n,1}(z)t^n
\end{array}\right).
\end{equation}
Hence, the elements $T$, $E^+$ and $E^-$ acts as
\begin{eqnarray}
TV_n&=&\left(\begin{array}{cc}
n+\dfrac{\alpha+1}{2} & 0 \\
0 & n+\dfrac{\alpha+1}{2}
\end{array}\right)V_n,\\
E^+V_n&=&\left(\begin{array}{cc}
n+1 & 0 \\
0 & n
\end{array}\right)V_{n+1},\\
E^-V_n&=&\left(\begin{array}{cc}
-n-\alpha & 0 \\
1 & -n-1-\alpha
\end{array}\right)V_{n-1}.
\end{eqnarray}
The Casimir can be represented as
\begin{equation}
\mathcal{K}V_n=\left(\begin{array}{cc}
(\alpha+5)(\alpha+3)/2 & 0 \\
n & (\alpha+3)(\alpha+1)/2
\end{array}\right)V_n.
\end{equation}

The polynomial $\omega_{n,1}(z)$ can be solved explicitly as a sum of Laguerre polynomials, i.e.

\begin{equation}
\omega_{n,1}(z)= \sigma_1 \left( (\alpha+1))\sum_{k=0}^{n-1}L_k^{(\alpha)}(z)  - n L_n^{(\alpha)}(z) \right)  - \sum_{k=0}^{n-1}L_k^{(\alpha)}(z).
\end{equation}

%\begin{equation}
%\omega_{n,1}(z)= - n \sigma_1L_n^{(\alpha)}(z)  - (1-\sigma_1(\alpha+1))\sum_{k=0}^{n-1}L_k^{(\alpha)}(z).
%\end{equation}

Using the lowest state on the second chain, i.e. $\vert 1,1\rangle=(\sigma_1z-1)t$, the generating function for the second chain takes the form
\begin{equation}
(\sigma_1z-1+t)t(1-t)^{-(\alpha+3)}\exp\left(\dfrac{-zt}{1-t}\right).
\end{equation}

\subsection{The three-chain example}
Let us define the vector
\begin{equation}
V_n=\left(\begin{array}{c}
\vert n,3\rangle\\
\\
\vert n,2\rangle \\
\\
\vert n,1\rangle
\end{array}\right)=\left(\begin{array}{c}
L_n^{(\alpha)}(z)t^n\\
\\
\omega_{n,1}(z)t^n\\
\\
\omega_{n,2}(z)t^n
\end{array}\right).
\end{equation}
Hence, the elements $T$, $E^+$ and $E^-$ acts as
\begin{eqnarray}
TV_n&=&\left(\begin{array}{ccc}
n+\dfrac{\alpha+1}{2} & 0 & 0\\
0 & n+\dfrac{\alpha+1}{2} & 0\\
0 & 0 & n+\dfrac{\alpha+1}{2}
\end{array}\right)V_n,\\
E^+V_n&=&\left(\begin{array}{ccc}
n+1 & 0 & 0 \\
0 & n & 0 \\
0 & 0 & n-1
\end{array}\right)V_{n+1},\\
E^-V_n&=&\left(\begin{array}{ccc}
-n-\alpha & 0 & 0 \\
1 & -n-1-\alpha & 0 \\
0 & 1 & -n-2-\alpha
\end{array}\right)V_{n-1}.
\end{eqnarray}
The Casimir can be represented as
\begin{equation}
\mathcal{K}V_n=\left(\begin{array}{ccc}
(\alpha+7)(\alpha+5)/2 & 0 & 0 \\
n-1 & (\alpha+5)(\alpha+3)/2 & 0 \\
0 & n & (\alpha+3)(\alpha+1)/2
\end{array}\right)V_n.
\end{equation}
An explicit expression for the polynomials is
\begin{equation}
\omega_{n,2}= n (n-1) \sigma_2 L_n^{(\alpha)}(z) + (n-1) (\sigma_1 -2 \sigma_2 (2+\alpha) ) L_{n-1}^{(\alpha)}(z) 
\end{equation}
\[+ \sum_{k=0}^{n-2}  (   (\frac{n-k-1}{2}) - \sigma_1 (  (n-k-1) \alpha + (n-1-2k)   )   +  \sigma_2 (2+\alpha) ( (n-k-1)\alpha + (n-1-3k)   )      )  L_{k}^{(\alpha)}(z). \]

Using the lowest state on the second chain, i.e. $\vert 2,1\rangle=(\sigma_2z^2-\sigma_1z+1/2)t^2$, the generating function for the second chain takes the form
\begin{equation}
(\sigma_2z^2-\sigma_1z(1-t)+(1-t)^2/2)t^2(1-t)^{-(\alpha+5)}\exp\left(\dfrac{-zt}{1-t}\right).
\end{equation}

\section{Properties of the polynomials}

In the construction of those indecomposable representations, there are various special functions, first at the level of the two variables differential operators satisfying an $sl(2)$ Lie algebra which involve the element of  $V_n$ that take the form of two-variables special functions. The $\omega_{n,i}$ are  building block of the representations for the diagonal and nondiagonal cases. Those $\omega_{n,i}$ also satisfy various differential-recurrence relations based on the raising and lowering operators $E^{+}$ and $E^{-}$. We have also provided generating functions and explicit solutions in terms of linear combinations of Laguerre orthogonal polynomials which allow to get further insight into their properties. This section will be devoted to provide ordinary differential equations, further recurrence relations and properties such as bi-orthogonality type relations with another set of related polynomial and zeroes of those polynomials in the case of the two chains.

\subsection{T nondiagonal}

It is interesting to note that the functions $\omega_{n,i}$ satisfy coupled homegeneous/non-homegenous equations chains as well as recurrence relations but also this differential equations which can be used to study the further properties:
\begin{equation} \label{recurdiff}
 \omega_{n,1}(z) - w'_{n,1}(z) + w'_{n+1,1}(z) =0.
 \end{equation} 
Another relation obtained directly from the recurrence relation is the following
\begin{equation}
  -(\alpha+n)w_{n-1,1}(z) +n w_{n,1}(z)     + (1+\alpha+n-z) w_{n,1} -(1+n) w_{n+1,1} =2 L_{n-1}^{\alpha}(z)-2 L_{n}^{\alpha}(z).
\end{equation}

Taking a sum of the first relation (\ref{recurdiff}) gives
\[  \omega_{k+1,1}'(z) + \sum_{i=0}^k \omega_{i,1}(z) =0. \]

Here we will provide some more properties in particular for $\omega_{n,1}$. Using the homogeneous differential equation for the Laguerre and the inhommogeneous differential equation for $\omega_{n,1}$, we see that they satisfy this fourth-order, homogeneous differential equation
\[ z^2 \omega_{n,1}'''' + ( (5+2\alpha)z -2 z^2) \omega_{n,1}''' + ( (2n-5 -2\alpha) z +z^2)\omega_{n,1}''\]
\[ + ( (n-1)(3+2\alpha)- (2n-2)z ) \omega_{n,1}' + n(n-1) \omega_{n,1} =0 \]

This equation display some similarities with other context such a multiple Laguerre or Hermite polynomial which are orthogonal. Here the polynomial are not orthogonal under the usual Laguerre measure but
satisfy analog of an bi-orthogonality relation of the form
\[  \int_0^{\infty}  q_{m}(z) \omega_{n,1}  x^{\alpha} e^{-z} dz =0 ,\quad \forall n < m, \]
where $q_{m}(z)$ are another set of polynomial which can be explicitly written as
\[ q_m(z)= \sum_{k=0}^n \binom{n}{k} (-1)^{2n+k} \frac{z^k}{(1+\alpha)_{k}} .\]

Note that for different choices of the parameters and $n$ the zeroes are on the real axis.

\begin{figure}[h!]
    \includegraphics[width=1.\textwidth]{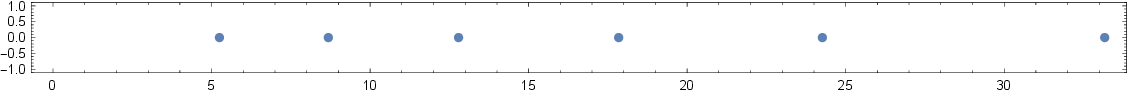}
   \caption{Zeroes of $\omega_{n,1}$ for $n=6$, $\sigma=2$ and $\alpha=10$  }   
\end{figure}

\begin{figure}[h!]
    \includegraphics[width=1.\textwidth]{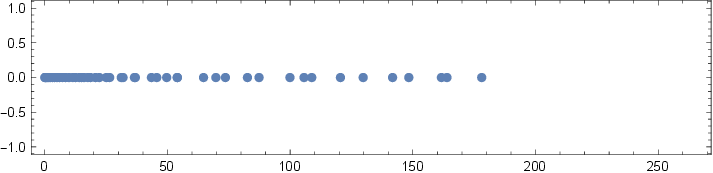}
   \caption{Zeroes of $\omega_{n,1}$ for $n=50$, $\sigma=\frac{1}{100}$ and $\alpha=\frac{1}{100}$  }   
\end{figure}

\subsection{T diagonal}

We consider the polynomials $\omega_{n,1}$ from the T diagonal related to Laguerre case. Those polynomial can be re-expressed as linear combination of Laguerre and possess generating function. They also satisfy as demonstrated various differential-recurrence relations which are direct consequences of the action of the generator of the sl(2) algebra. We also have the following relation for given $\omega_{n,i}$
\[  (n+\alpha  +1 -z+n) \omega_{n,i}  = (n-i+1) \omega_{n+1,i} + (\alpha +n + i) \omega_{n-1,i}- \omega_{n-1,i-1}.  \]

In the case of $\omega_{n,1}$, it can be shown that
\[  ( -1 +\sigma_1 + \alpha \sigma_1  ) z \omega_{n,1}''' + (  (3+\alpha)(-1+ \sigma_1 + \sigma_1 \alpha) + (1 - \sigma_1 (-(n+1)+\alpha)  )z)  \omega_{n,1}''\]
\[ + ( -(n-1) + (2n-2) \sigma_1(1+\alpha) - n \sigma_1 z  ) \omega_{n,1}' + n^2 \sigma_1 \omega_{n,1} =0. \]

Other aspect such as orthogonality/bi-orthogonality can be investigated here. For example, we have the following orthogonality 
\[  \int_0^{\infty}  q_{m}(z) \omega_{n,1}  x^{\alpha} e^{-z} dz =0 ,\quad \forall n \leq m, \]
where
\[ q_m(z)= \sum_{k=0}^n \binom{n}{k} (-1)^{2n+k} \frac{z^k  (-1+ \sigma_1 + \alpha \sigma_1 ) }{(1+\alpha)_{k} (-1 +(k+1) \sigma_1 + \alpha \sigma_1) }. \]

For different choices of the parameters and $n$ the zeroes are on the real axis, as in  the nondiagonal case.

\begin{figure}[h!]
    \includegraphics[width=1.\textwidth]{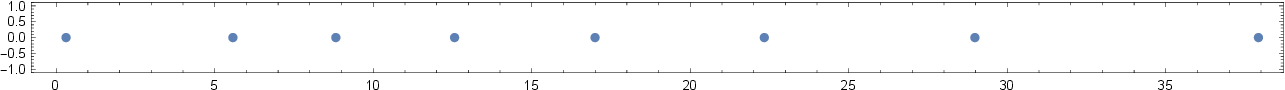}
   \caption{Zeroes of $\omega_{n,1}$ for $n=8$, $\sigma=2$ and $\alpha=10$  }   
\end{figure}
\begin{figure}[h!]
    \includegraphics[width=1.\textwidth]{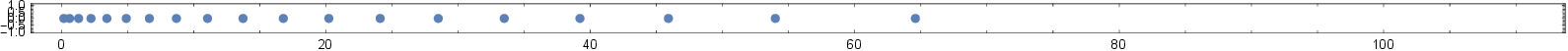}
   \caption{Zeroes of $\omega_{n,1}$ for $n=20$, $\sigma=\frac{1}{1000}$ and $\alpha=\frac{1}{1000}$  }   
\end{figure}

\section{Further algebraic aspects of the solutions}

We have described two cases of representations, with T diagonalizable and T nondiagonalible (both in context of Laguerre polynomials). We have obtained analogs of the Rodrigues formula from (3.9) and (4.6), which generalized well known formula for classical orthogonal polynomials.

In those case of $sl(2,\mathbb{C})$ indecomposable representations we can understand the construction from point of view of induced representation and the modules can be generated from an n-dimensional indecomposable representation of the Borel of $sl(2,\mathbb{C})$. The those modules can be related via the following transformation of generators (taking $E^{-}=-e$ and $E^{+}=-f$) and by multiplicative factor of the module:

%\newline
%\newline

\begin{equation}
[e,f]=h,
\end{equation}
\begin{equation}
[h,e]=2e,
\end{equation}
\begin{equation}
[h,f]=-2f,
\end{equation}
with further relations
\[ [h,e^{a}]=2a e^{a}, \]
\[ [f,e^{a}]=-a(h+a-1)e^{a-1},\]
\[ [e,f^{a}]=a(h+a-1)f^{a-1}, \]
\[ [h,f^{a}]=-2 a f^{a}. \]

\subsection*{T nondiagonalizable}

We consider first the case where the generator of $sl(2)$ is nondiagonalizable. The modules which form the representations of the Borel subalgebra $\{e,h\}$ can be denoted $v_{(1)0}^{(n)}$ and then satisfy the explicit action

\begin{eqnarray}
&& h v_{(1)0}^{(n)}=\lambda v_{(1)0}^{(n)},  \\
&& h v_{(1)0}^{(k)}=\lambda v_{(1)0}^{(k)}+ v_{(1)0}^{(k+1)},\quad k=1,...,n-1, \nonumber \\
&& e v_{(1)0}^{(i)} =0,\quad i=1,...,n. \nonumber
\end{eqnarray}
Here the index $(1)$ refer to the choice of representations for the Borel subalgebra $v_{(1)0}^{(n)}$ which can be seen as the choice of a column ( $n \times 1$) in the space of $v_{(i)0}^{(j)}$ which consist of a lattice ($n \times N$) formed by the collection of generalized Weight spaces. They are related via the generator $e$. Here we have one of such generalized Weight space of dimension $n$. 

Then generating the other chains in the following way with modules denoted by $v_{(1)a}^{(i)}$ for $i=1,..,n-1$  can be done as follow:

\begin{eqnarray}
&& v_{(1)a}^{(i)}=f^{a} v_{(1)0}^{(i)}, i=1,..,n-1,  \\
&& v_{(1)a}^{(n)}=f^{a} v_{(1)0}^{(n)}.  \nonumber
\end{eqnarray}

Using commutation relations we obtain the following action

\begin{eqnarray}
&& h v_{(1)a}^{(i)} = (\lambda -2 a) v_{(1)a}^{(i)} + v_{(1)a}^{(i+1)},  \\
&& h v_{(1)a}^{(n)} =(\lambda-2 a) v_{(1)a}^{(n)},                \nonumber \\
&& f v_{(1)a}^{(i)} = v_{(1)a+1}^{(i)}, \quad i=1,...,n-1,                 \nonumber \\
&& f v_{(1)a}^{(n)} = v_{(1)a+1}^{(n)},                 \nonumber \\
&& e v_{(1)a}^{(i)} =a (\lambda-a+1) v_{(1)a-1}^{(i)} + a v_{(1)a-1}^{(i+1)},                 \nonumber \\
&& e v_{(1)a}^{(n)} = a (\lambda -a +1) v_{(1)a-1}^{(n)}.                   \nonumber
\end{eqnarray}

%The those modules can be related via the following transformation of generators (taking $T^{-}=-e$ and $T=-h$) and by multiplicative factor of the module.

%\newline
%\newline

\subsection*{T diagonalizable} 

We now consider the case where ($T$ i.e. the Cartan element h is diagonalizable ). The modules forming the representation of the Borel subalgebra ($\{e,h\}$ or $\{E^{-},T\}$) are given by $v_{(k)0}^{(1)}$,

\begin{eqnarray}
&&  h v_{(k)0}^{(1)}=(\lambda +2 (k-1)) v_{(k)0}^{(1)},  \\
&&  e v_{(n)0}^{(1)} =0, \nonumber \\
&&  e v_{(i)0}^{(1)}=v_{(i+1)0}^{(1)},\quad i=1,...,n-1, \nonumber 
\end{eqnarray}

Here the index $(1)$ refer to the choice of representations for the Borel subalgebra $v_{(i)0}^{(1)}$ which can be seen as the choice of a row in the space of $v_{(i)0}^{(j)}$ which consist of a lattice ($n \times N$) formed by collections of generalized Weight spaces (indexed by $i$). Here we have $n$ one-dimensional Weight spaces. 

with the modules of the other chains are constructed via 
\begin{eqnarray}
&&  v_{(i)a}^{(1)}=f^{a} v_{(i)0}^{(1)},  \\
&&    v_{(n)a}^{(1)}=f^{a} v_{(n)0}^{(1)}, \nonumber 
\end{eqnarray}
and we have
\begin{eqnarray}
&&   h v_{(i)a}^{(1)}=(\lambda +2(i-1)-2a) v_{(i)a}^{(1)},  \\
&&    h v_{(n)a}^{(1)}=(\lambda +2(n-1)-2a) v_{(n)a}^{(1)}  ,\nonumber \\
&&    f v_{(n)a}^{(1)}=v_{(n)a+1}^{(1)},             \nonumber \\
&&    f v_{(i)a}^{(1)}=v_{(i)a+1}^{(1)} , \quad i=1,...n-1,               \nonumber \\
&&     e v_{(i)a}^{(1)}=a(\lambda+2i -a-1) v_{(i)a-1}^{(1)} + v_{(i+1)a}^{(1)},               \nonumber \\
&&     e v_{(n)a}^{(1)}=a(\lambda+2 n -a -1) v_{(n)a-1}^{(1)}.               \nonumber
\end{eqnarray}

Realization of these representations as differential operators has been a so far unexplored subject and the results above represent the first steps of a program in which hypergeometric, Laguerre and other special functions as well as generalization in term of non homogeneous linear differential equations are considered in this context.

\section{Conclusion}

In this paper we have presented a complete description of two different types of infinite dimensional indecomposable representations of $sl(2)$ involving N chains and their related differential operator realizations with two variables. They correspond to two cases for which the action of the generator T is diagonalizable or nondiagonalizable. Those results constitute generalization of problems considered in \cite{mil68,vil91} previously for $sl(2)$. \par
%%%%%%
We have presented several results such as second-order differential-recurrence relations for the $\omega_{n,i}$ which are polynomials that are part of the chain as building blocks. Those polynomials also satisfy for the polynomials $\omega_{n,1}$ of T diagonal case and T nondiagonal case respectively third- and fourth-order ordinary differential equations. We have also presented further recurrence formula, generating functions and related polynomials satisfying bi-orthogonal type of relations. We have also investigated zeroes of those new polynomials and they lie on a segment on the real axis. Let us remark that those polynomials are not orthogonal under Laguerre's weight, and there are many instances of polynomials beyond classical orthogonal polynomials that are of interest for applications in physics and in particular in quantum physics. Among them the para-Jacobi polynomials that was exploited in context of supersymmetric quantum mechanics. We can think of Bender-Dunne polynomial referred as quasi-orthogonal polynomials. \par
%%%%%%
From a perspective of systems of coupled non-homogeneous ODEs the systems considered in this papers extend those previously introduced \cite{bab67,mil70a,zac76, bac86,bac01} and proposed an algebraic setting. As special functions and in particular Laguerre appear in many context and in particular related to solution of different PDEs among them the Schr\"odinger equations and as solution of ODE obtained from separation of variables \cite{kal86,mil84} this is other context where results may find useful applications. \par
%%%%
The case of non-homogeneous systems of ODE as also been seen in context of supersymmetric quantum mechanics and was involving Jordan blocks \cite{sch06,con15}. Other areas such as pseudo-hermitian Hamiltonian were also known to be connected with representations involving Jordan blocks \cite{mar22a,mar22b}. The results obtained in this paper and further generalization are likely to find application in context of quantum mechanical problem or other setting of mathematical physics.

\section*{Data availability statement}

The data that support the findings of this study are available upon reasonable request from the
authors.

\section*{Acknowledgement}

IM was supported by the Australian Research Council Future Fellowship FT180100099.

\end{document}